\documentclass{article}

\usepackage{amsfonts}
\usepackage{amsmath}

\title{Gradings on $o(8,\mathbb C)$}
\author{Cristina Draper \thanks{\ Supported by MCYT
grants MTM2004-06580-C02-02 and MTM2004-08115-C04-04, and by JA
grants FQM-336 and FQM-1215}\\ Departamento de Matem\'{a}tica
Aplicada, Universidad de M\'{a}laga
\\ cdf@uma.es\\[1ex]
 Antonio Viruel\thanks{\ Partially supported by the MCYT
grant MTM2007-60016, and by the JA grant FQM-213}\\Departamento de
\'{A}lgebra, Geometr\'{\i}a y Topolog\'{\i}a, Universidad de
M\'{a}laga\\viruel@agt.cie.uma.es }

\begin{document}

\maketitle
\begin{abstract}
      This is a matricial description of all the fine group gradings on
      the exceptional Lie algebra
      $o(8,\mathbb C)$. There are fourteen.
\end{abstract}


\section{Introduction}

Given a Lie algebra $L$, a grading on $L$  is a decomposition of
vector spaces $L=\oplus_{i\in I}L_i$ such that $L_i\ne0$ and for
each pair of indices $i,j\in I$ there exists $k\in I$ such that
$L_iL_j\subset L_k$ (see Definition 2.1 below). Besides of its own
mathematical interest, gradings on Lie algebras show up in
numerous applications of physical nature.

As it was already pointed out by Patera et al.\ in a series of
papers on the subject \cite{pat0, pat1, pat2, pat3, pat4},
gradings are in the background of the choice of bases and the
additive quantum numbers, and are a key ingredient when studying
contractions that keep ``undeformed" certain subalgebra (e.g. see
\cite{klim}).

Gradings on Lie algebras also play a relevant role in quantum
mechanics via Jordan algebras \cite{jordan}: given a pair of
indices $i,j\in I$ verifying $(L_iL_j)L_i\subset L_i$, any element
$k\in L_j$ gives rise to a Jordan algebra structure in $L_i$ by
defining $x\circ y=(xk)y$.

Gradings on the algebras $o(n,\mathbb C)$ have been computed in
\cite{Ivan2} and \cite{LGII} for $n\ne8$. The case $n=8$ is of
special relevance because of its strong symmetry, reflected partly
by the triality automorphism. This makes that our matrices algebra
is usually considered as a exceptional Lie algebra. Some works
with a treatment of gradings on exceptional Lie algebras are
\cite{g2,f4}. Here we describe in detail all the fourteen fine
gradings on $o(8,\mathbb C)$ (see Section 3). The proof showing
that these are all the fine gradings is lengthy and cannot be
included here, so it will appear in \cite{d4}.

\section{Basic concepts}

In this minimal section, we introduce the basic concepts we use
along this manuscript.

 {\bf Definition 2.1.} \quad
  A  \emph{grading}  on a Lie algebra $L$ over a group $G$  is a decomposition in vector spaces
  $L=\oplus_{g\in G}L_g$ such that  $L_gL_h\subset L_{gh}$ for all $g,h\in G$ (sometimes called group grading).
   It will also be required that $G$
  is generated by $\{g\in G\mid L_g\ne0\}$. We say that a grading is of
\emph{type} $(h_1,\dots,h_s)$ when there are $h_i$ components of
dimension $i$, and $s$ is the greatest nonzero dimension of a
component.\smallskip

   {\bf Definition 2.2.} \quad
   A grading  is called \emph{fine} when it cannot be refined. A \emph{refinement} of the grading  $L=\oplus_{g\in G}L_g$
   is another (group) grading $L=\oplus_{h\in H}M_h$ such that each component $L_g$ is sum of several components
   $M_h$.\smallskip

{\bf Example 2.1.} \quad
   The main example of a fine grading is the root decomposition in a simple Lie algebra, which turns out to be a grading
   over $\mathbb Z^l$, for $l$ the rank of the algebra.\smallskip

From now on, $L$ will denote the algebra of skewsymmetric matrices
$o(8,\mathbb C)=\{x\in Mat_{8\times8}(\mathbb C)\mid x+x^t=0\}$.
Let $e_{ij}$ be the elementary matrix   with $(i,j)$-entry $1$,
and the remaining entries zero, and define $b_{ij}=e_{ji}-e_{ij}$.
Then $\{b_{ij}\mid i,j=1,\dots 8,i<j\}$ is a basis of $L$, and the
following section describes   the fine gradings on $L$ in terms of
this base.

Recall that a grading on $L$ is always the simultaneous
diagonalization relative to a commuting set of semisimple
automorphisms (more details in \cite{g2}). The grading is fine
when it is produced by a maximal subgroup of semisimple
automorphisms, which is usually called a \emph{MAD}.

\section{The fine gradings}
We introduce some useful elements of $O(8,\mathbb C)$ to describe
the gradings on $L$.

{\small $$
 \begin{array}{rl}
 g(a)= & e_{11 }+e_{ 22}+e_{33 }+e_{44
}+ \frac12(a+\frac1a) (e_{55 }+e_{66 }+e_{77 }+e_{ 88})+ \frac
i2(a-\frac1a)(e_{57 }+e_{68 }+e_{75 }+e_{86 })\\
f(a)=& \frac12(a+\frac1a)(  e_{11 }+e_{ 22}+e_{33 }+e_{44 })+
 (e_{55 }+e_{66 }+e_{77 }+e_{ 88}) +\frac
i2(a-\frac1a)(e_{13 }+e_{24 }+e_{31 }+e_{42 })\\
h(a)=& \frac12(a+\frac1a)(  e_{11 }+e_{ 22}+e_{33 }+e_{44 }+
 e_{55 }+e_{66 }+e_{77 }+e_{ 88}) \\&+\frac
i2(a-\frac1a)( e_{ 15} +e_{ 26}+e_{37 }+e_{48 }+e_{ 51}+e_{62
}+e_{ 73}+e_{ 84})\\
 p(a) =& e_{11 }+e_{ 22}+e_{33 }+e_{44 }+
 e_{55 }+e_{66 } +\frac12(a+\frac1a)(e_{77 }+e_{ 88})+ \frac
i2(a-\frac1a)( e_{78 }+e_{87 })\\
 q(a) =& e_{11 }+e_{ 22}+e_{33
}+e_{44 } +\frac12(a+\frac1a)(e_{55 }+e_{ 66})+ \frac
i2(a-\frac1a)( e_{56 }+e_{65 })+
 e_{77 }+e_{88 } \\
 r(a) =& e_{11 }+e_{ 22}+\frac12(a+\frac1a)(e_{33 }+e_{ 44})+ \frac
i2(a-\frac1a)( e_{34 }+e_{43 })+e_{55 }+e_{66 }+
 e_{77 }+e_{88 }\\
 s(a) =& \frac12(a+\frac1a)(e_{11 }+e_{ 22})+ \frac
i2(a-\frac1a)( e_{12 }+e_{21 })+e_{33 }+e_{ 44}+e_{55 }+e_{66 }+
 e_{77 }+e_{88 }\end{array} $$

$$\begin{array}{ll}f_1=\textrm{diag}\{1,1,1,1,1,1,1,-1\} &
   f_2=\textrm{diag}\{1,1,1,1,1,1,-1,1\}\\
   f_3=\textrm{diag}\{1,1,1,1,1,-1,1,1\}&
   f_4=\textrm{diag}\{1,1,1,1,-1,1,1,1\}
   \\
   f_5=\textrm{diag}\{1,1,1,-1,1,1,1,1\}&
   f_6=\textrm{diag}\{1,1,-1,1,1,1,1,1\}\\
   f_7=\textrm{diag}\{1,-1,1,1,1,1,1,1\}&
   f_8=\textrm{diag}\{-1,1,1,1,1,1,1,1\}
      \end{array}$$
$$\begin{array}{lllll} g_1=f_8f_7&
    g_2=f_6f_5 &g_4=f_7f_5f_4f_2
    &g_5=f_8f_6f_4f_2&
   \\
    g_6=f_2f_1 &g_7=f_7f_5f_3f_1 &g_9=f_8f_7f_4f_3&g_{10}=f_6f_5f_2f_1& g_{13}=f_4f_3f_2f_1 \end{array}$$

 $$\begin{array}{rl}g_3 =& e_{ 12}+e_{21 }+e_{34
}+e_{43 }+e_{ 56}+e_{ 65}+e_{ 78}+e_{
 87}\\
  g_8 = & e_{13 }+e_{31 }+e_{ 24}+e_{42 }+e_{57 }+e_{75 }+e_{ 68}+e_{86
  }\\
   g_{11} =& e_{ 12}+e_{21 }+e_{34 }+e_{ 43}-e_{55 }+e_{ 66}-e_{ 77}+e_{ 88}
   \\
    g_{12} = &e_{11 }-e_{ 22}+e_{33 }-e_{44 }-e_{ 56}+e_{65 }-e_{ 78}+e_{87
    }\\
    g_{14}=&e_{ 15}+e_{ 26}+e_{37 }+e_{48 }+e_{51 }+e_{62 }+e_{ 73}+e_{84
    }
   \end{array} $$}

Notice that $f_i,g_i,f(a),g(a),h(a),p(a),q(a),r(a),s(a)\in
O(8,\mathbb C)=\{P\in Mat_{8\times8}(\mathbb C)\mid
PP^t=\hbox{id}\}$ for all $a\in\mathbb C^*$. Thus we can use the
adjoint map to obtain automorphisms of $L$ by means of
$\mathop{\rm Ad}\colon O(8,\mathbb C)\to \mathop{\rm Aut}
o(8,\mathbb C)$, $\mathop{\rm Ad} P(x)=P^{-1}xP$. In particular,
denote $F_i=\mathop{\rm Ad} f_i$ and $G_i=\mathop{\rm Ad} g_i$.
With all these elements, we are prepared to describe completely
the group of automorphisms producing each grading, denoted by
$Q_i$. We also provide the simultaneous diagonalization,  which is
a straightforward computation by means of any computer algebra
software.\smallskip

   \textbf{Grading over $\mathbb Z\times \mathbb Z_2^4$:}
$Q_1=\langle\{\mathop{\rm Ad} g(u), G_1,G_2,G_3,G_4\mid
u\in\mathbb C^*\}\rangle$
  is an abelian diagonalizable subgroup of $\mathop{\rm Aut} L$. In fact it is
  maximal   according to \cite{LGII} (it is the MAD corresponding to $T_{2,2}^{(1)}$).
  Hence it produces a fine grading, of type $(25,0,1)$, whose homogeneous
  components are:

  {\small $$\begin{array}{ll}
  L_{\frac12,1,-1,-1,-1}=\langle ib_{3,5}-b_{3,7}-ib_{4,6}+b_{4,8}\}\rangle&L_{\frac14,1,1,-1,-1}=\langle -b_{5,6}-ib_{5,8}+ib_{6,7}+b_{7,8}  \rangle\\
 L_{\frac12,1,-1,-1,1}=\langle ib_{3,6}-b_{3,8}-ib_{4,5}+b_{4,7}  \rangle&L_{\frac12,-1,1,-1,-1}=\langle ib_{1,5}-b_{1,7}-ib_{2,6}+b_{2,8}  \rangle\\
  L_{\frac12,1,-1,1,-1}=\langle -ib_{3,5}+b_{3,7}-ib_{4,6}+b_{4,8}  \rangle&L_{\frac12,-1,1,-1,1}=\langle ib_{1,6}-b_{1,8}-ib_{2,5}+b_{2,7}  \rangle\\
   L_{\frac12,1,-1,1,1}=\langle -ib_{3,6}+b_{3,8}-ib_{4,5}+b_{4,7}  \rangle&L_{\frac12,-1,1,1,-1}=\langle -ib_{1,5}+b_{1,7}-ib_{2,6}+b_{2,8}  \rangle\\
     L_{\frac12,-1,1,1,1}=\langle -ib_{1,6}+b_{1,8}-ib_{2,5}+b_{2,7}  \rangle&
     L_{1,1,1,-1,-1}=\langle b_{1,2},\,b_{3,4},\,b_{7,8}+b_{5,6}  \rangle\\
         L_{2,1,-1,-1,1}=\langle -ib_{3,6}-b_{3,8}+ib_{4,5}+b_{4,7}  \rangle&L_{2,-1,1,-1,-1}=\langle -ib_{1,5}-b_{1,7}+ib_{2,6}+b_{2,8}  \rangle\\
          L_{2,1,-1,1,-1}=\langle ib_{3,5}+b_{3,7}+ib_{4,6}+b_{4,8}  \rangle&L_{2,-1,1,-1,1}=\langle -ib_{1,6}-b_{1,8}+ib_{2,5}+b_{2,7}  \rangle\\
           L_{2,1,-1,1,1}=\langle ib_{3,6}+b_{3,8}+ib_{4,5}+b_{4,7}  \rangle&L_{2,-1,1,1,-1}=\langle ib_{1,5}+b_{1,7}+ib_{2,6}+b_{2,8}  \rangle\\
            L_{4,1,1,-1,-1}=\langle -b_{5,6}+ib_{5,8}-ib_{6,7}+b_{7,8}  \rangle&L_{2,-1,1,1,1}=\langle ib_{1,6}+b_{1,8}+ib_{2,5}+b_{2,7}  \rangle\\
            L_{1,1,1,1,-1}=\langle b_{5,8}+b_{6,7}  \rangle&
L_{2,1,-1,-1,-1}=\langle -ib_{3,5}-b_{3,7}+ib_{4,6}+b_{4,8}
\rangle\end{array}$$
$$\begin{array}{lll}L_{1,-1,-1,-1,-1}=\langle b_{2,3}-b_{1,4}  \rangle&L_{1,1,1,1,1}=\langle b_{6,8}+b_{5,7}  \rangle&
L_{1,-1,-1,1,-1}=\langle b_{2,3}+b_{1,4}  \rangle\\
L_{1,-1,-1,-1,1}=\langle b_{2,4}-b_{1,3}
\rangle&L_{1,1,1,-1,1}=\langle b_{6,8}-b_{5,7}  \rangle&
L_{1,-1,-1,1,1}=\langle b_{2,4}+b_{1,3}  \rangle\end{array}$$ }

 The indices of the components are not the indices of
the grading, but the eigenvalues that $\{\mathop{\rm Ad} g(2),
G_1,G_2,G_3,G_4\}$ act with (e.g. the index
$\{\frac12,1,-1,-1,-1\}$ is for the element
$(-2,\bar0,\bar1,\bar1,\bar1)\in\mathbb Z\times\mathbb Z_2^4$).
  The universal group of the grading (see
\cite{g2} for the definition) is just $\mathbb Z\times \mathbb
Z_2^4(=\chi(Q_1))$. These comments about the universal group and
the order and eigenvalues of the automorphisms, are valid for all
the gradings in this paper.
\smallskip

\textbf{Grading over $\mathbb Z^2\times \mathbb Z_2^2$:}
$Q_2=\langle\{\mathop{\rm Ad} f(u), \mathop{\rm Ad} g(v),
G_3,G_5\mid u,v\in\mathbb C^*\}\rangle\cong(\mathbb
C^*)^2\times\mathbb Z_2^2$
  is an abelian diagonalizable subgroup of $\mathop{\rm Aut} L$. In fact, it
  is the MAD corresponding to $T_{ 4,0}^{( 1)}$ (\cite{LGII}).
  The simultaneous diagonalization of $L$ relative to this subgroup
  of automorphisms is of type $(20,4)$:
{\small $$\begin{array}{l}L_{ \frac{1}{2}, \frac{1}{3}, -1, -1
}=\langle  b_{1, 6} + i b_{1, 8} - b_{2, 5} - i b_{2, 7}+ i b_{3,
6} -b_{3, 8} - i b_{4, 5} + b_{4, 7} \rangle\\
L_{\frac{1}{2}, \frac{1}{3}, -1, 1 }=\langle   b_{1, 5} + i b_{1,
7} -
    b_{2, 6} - i b_{2, 8} + i b_{3, 5} -
    b_{3, 7} - i b_{4, 6} + b_{4, 8} \rangle
    \\L_{ \frac{1}{2}, \frac{1}{3}, 1, -1
}=\langle   -b_{1, 6} - i b_{1, 8} -
    b_{2, 5} - i b_{2, 7} - i b_{3, 6} +
    b_{3, 8} - i b_{4, 5} + b_{4, 7} \rangle\end{array}$$}

{\small $$\begin{array}{l}
    L_{ \frac{1}{2}, \frac{1}{3}, 1, 1
}=\langle  -b_{1, 5} - i b_{1, 7} -
    b_{2, 6} - i b_{2, 8} - i b_{3, 5} +
    b_{3, 7}- i b_{4, 6} + b_{4, 8} \rangle\\L_{  \frac{1}{2}, 3, -1, -1
}=\langle  -b_{1, 6} + i b_{1, 8} +
    b_{2, 5} - i b_{2, 7} - i b_{3, 6} -
    b_{3, 8}+ i b_{4, 5} + b_{4, 7}   \rangle\\
L_{ \frac{1}{2}, 3, -1, 1 }=\langle  -b_{1, 5} + i b_{1, 7} +
    b_{2, 6} - i b_{2, 8} - i b_{3, 5} -
    b_{3, 7}+ i b_{4, 6} + b_{4, 8}  \rangle\\
L_{ \frac{1}{2}, 3, 1, -1 }=\langle  b_{1, 6} - i b_{1, 8} +
    b_{2, 5} - i b_{2, 7} + i b_{3, 6} +
    b_{3, 8}+ i b_{4, 5} + b_{4, 7}  \rangle\\
L_{ \frac{1}{2}, 3, 1, 1 }=\langle  b_{1, 5} - i b_{1, 7} +
    b_{2, 6} - i b_{2, 8} + i b_{3, 5} +
    b_{3, 7}+ i b_{4, 6} + b_{4, 8}  \rangle\\
L_{ 2, \frac{1}{3}, -1, -1 }=\langle  -b_{1, 6} - i b_{1, 8} +
    b_{2, 5} + i b_{2, 7} + i b_{3, 6} -
    b_{3, 8}- i b_{4, 5} + b_{4, 7}  \rangle\\
L_{  2, \frac{1}{3}, -1, 1 }=\langle  -b_{1, 5} - i b_{1, 7} +
    b_{2, 6} + i b_{2, 8} + i b_{3, 5} -
    b_{3, 7}- i b_{4, 6} + b_{4, 8}  \rangle\\
L_{ 2, \frac{1}{3}, 1, -1 }=\langle  b_{1, 6} + i b_{1, 8} +
    b_{2, 5} + i b_{2, 7} - i b_{3, 6} +
    b_{3, 8}- i b_{4, 5} + b_{4, 7}  \rangle\\
L_{ 2, \frac{1}{3}, 1, 1 }=\langle  b_{1, 5} + i b_{1, 7} +
    b_{2, 6} + i b_{2, 8} - i b_{3, 5} +
    b_{3, 7}- i b_{4, 6} + b_{4, 8}  \rangle\\
L_{ 2, 3, -1, -1 }=\langle  b_{1, 6} - i b_{1, 8} -
    b_{2, 5} + i b_{2, 7} - i b_{3, 6} -
    b_{3, 8}+ i b_{4, 5} + b_{4, 7}  \rangle\\
L_{  2, 3, -1, 1 }=\langle  b_{1, 5} - i b_{1, 7} -
    b_{2, 6} + i b_{2, 8} - i b_{3, 5} -
    b_{3, 7}+ i b_{4, 6} + b_{4, 8}  \rangle\\
L_{ 2, 3, 1, -1 }=\langle  -b_{1, 6} + i b_{1, 8} -
    b_{2, 5} + i b_{2, 7} + i b_{3, 6} +
    b_{3, 8}+ i b_{4, 5} + b_{4, 7}  \rangle\\
L_{ 2, 3, 1, 1 }=\langle  -b_{1, 5} + i b_{1, 7} -
    b_{2, 6} + i b_{2, 8} + i b_{3, 5} +
    b_{3, 7}+ i b_{4, 6} + b_{4, 8}  \rangle\\
\end{array}$$} {\small $$\begin{array}{ll}
L_{ 1, 1, 1, 1 }=\langle  b_{1, 3} + b_{2, 4},\,b_{5, 7} + b_{6,
8} \rangle&
L_{ 1, 1, -1, -1 }=\langle   b_{1, 2} + b_{3, 4},\,b_{5, 6} + b_{7, 8}  \rangle\\
L_{ 1, 1, -1, 1 }=\langle  -b_{1, 3} + b_{2, 4},\,-b_{5, 7} +
b_{6, 8}  \rangle&
L_{ 1, 1, 1, -1 }=\langle  b_{1, 4} + b_{2, 3},\,b_{5, 8} + b_{6, 7}  \rangle\\
L_{\frac{1}{4}, 1, -1, -1}=\langle  -b_{1, 2} - i b_{1, 4} + i
b_{2, 3}+ b_{3, 4}  \rangle& L_{  1, 9, -1, -1 }=\langle  -b_{5,
6} + i b_{5, 8} - i b_{6, 7} + b_{7, 8} \rangle\\L_{ 4, 1, -1, -1
}=\langle  -b_{1, 2} + i b_{1, 4} - i b_{2, 3}+ b_{3, 4}
\rangle&L_{ 1,\frac{1}{9}, -1, -1 }=\langle -b_{5, 6} - i b_{5, 8}
+ i b_{6, 7} + b_{7, 8} \rangle\\\end{array}$$}

\textbf{Grading over $  \mathbb Z_2^5$:}
$Q_3=\langle\{G_6,G_1,G_2,G_3,G_7\}\rangle\cong\mathbb Z_2^5$
  is a MAD of $\mathop{\rm Aut} L$, corresponding to $T_{ 0,4}^{( 1)}$.
The simultaneous diagonalization of $L$ relative to the quasitorus
$Q_3$ is of type $(24,0,0,1)$, with $
      L_{ 1, 1, 1, -1, -1  }=\langle b_{ 1,2 },b_{ 3,4 },b_{ 5,6 },b_{ 7,8 } \rangle
      $ and: {\small $$\begin{array}{lll}
L_{ -1, -1, 1, -1, -1}=\langle -b_{ 1,8 }+b_{2 ,7 } \rangle & L_{
-1, -1, 1, -1, 1}=\langle -b_{ 1,7 }+b_{2 ,8 } \rangle & L_{ -1,
-1, 1,
      1, -1}=\langle b_{ 1,8 }+b_{2 ,7 } \rangle \\
      L_{ -1, -1, 1, 1, 1}=\langle b_{ 1,7 }+b_{2 ,8 } \rangle
      & L_{ -1, 1, -1, -1, -1}=\langle -b_{ 3,8 }+b_{4 ,7 } \rangle
      & L_{ -1,
      1, -1, -1, 1}=\langle -b_{ 3,7 }+b_{4 ,8 } \rangle
      \\ L_{ -1, 1, -1, 1, -1}=\langle b_{3 ,8 }+b_{4 ,7 } \rangle
      & L_{ -1, 1, -1, 1, 1}=\langle b_{ 3,7 }+b_{4 ,8 } \rangle
      & L_{ -1,
      1, 1, -1, -1}=\langle -b_{ 5,8 }+b_{6 ,7 } \rangle \\
      L_{ -1, 1, 1, -1, 1}=\langle -b_{ 5,7 }+b_{6 ,8 } \rangle &
       L_{ -1, 1, 1, 1, -1}=\langle b_{ 5,8 }+b_{6 ,7 } \rangle &
       L_{ -1, 1,
       1, 1, 1}=\langle b_{ 5,7 }+b_{ 6,8 } \rangle \\
       L_{ 1, -1, -1, -1, -1}=\langle -b_{ 1,4 }+b_{ 2,3 } \rangle &
       L_{ 1, -1, -1, -1, 1}=\langle -b_{ 1,3 }+b_{ 2,4 } \rangle &
       L_{
      1, -1, -1, 1, -1}=\langle b_{ 1,4 }+b_{ 2,3 } \rangle \\
      L_{ 1, -1, -1, 1, 1}=\langle b_{ 1,3 }+b_{ 2,4 } \rangle &
      L_{ 1, -1, 1, -1, -1}=\langle -b_{ 1,6 }+b_{ 2,5 } \rangle &
      L_{
      1, -1, 1, -1, 1}=\langle -b_{ 1,5 }+b_{ 2,6 } \rangle \\
      L_{ 1, -1, 1, 1, -1}=\langle b_{ 1,6 }+b_{ 2,5 } \rangle &
      L_{ 1, -1, 1, 1, 1}=\langle b_{ 1,5 }+b_{ 2,6 } \rangle &
      L_{ 1,
      1, -1, -1, -1}=\langle -b_{ 3,6 }+b_{ 4,5 } \rangle \\
      L_{ 1, 1, -1, -1, 1}=\langle -b_{ 3,5 }+b_{ 4,6 } \rangle &
      L_{ 1, 1, -1, 1, -1}=\langle b_{ 3,6 }+b_{ 4,5 } \rangle & L_{ 1,
      1, -1, 1, 1}=\langle b_{ 3,5 }+b_{ 4,6 } \rangle
\end{array}$$}

\textbf{Grading over $\mathbb Z\times \mathbb Z_2^4$:}
$Q_4=\langle\{\mathop{\rm Ad} h(u),G_8,G_9,G_3,G_5\mid u\in\mathbb
C^*\}\rangle\cong(\mathbb C^*)\times\mathbb Z_2^4$
  is a MAD of $\mathop{\rm Aut} L$, corresponding to $T_{ 2,0}^{( 2)}$.
The simultaneous diagonalization of $L$ relative to the quasitorus
$Q_4$ is of type $(28)$: {\small $$\begin{array}{ll}
      L_{ 1, -1, -1, -1, -1}=\langle  b_{1,8} - b_{2,7}- b_{3,6} + b_{4,5} \rangle&
      L_{ 1, -1, -1, -1, 1}=\langle -b_{1,3} + b_{2,4}- b_{5,7} + b_{6,8} \rangle\\
      L_{
      1, -1, -1, 1, -1}=\langle b_{1,4} + b_{2,3}+ b_{5,8} + b_{6,7} \rangle&
      L_{ 1, -1, -1, 1, 1}=\langle b_{1,3} + b_{2,4}+ b_{5,7} + b_{6,8} \rangle\\
      L_{ 1, -1, 1, -1, -1}=\langle -b_{1,2} + b_{3,4}- b_{5,6} + b_{7,8} \rangle&
      L_{
      1, -1, 1, -1, 1}=\langle b_{1,5} - b_{2,6}- b_{3,7} + b_{4,8} \rangle\\
      L_{ 1, -1, 1, 1, -1}=\langle -b_{1,6} - b_{2,5}+ b_{3,8} + b_{4,7} \rangle&
      L_{ 1, -1, 1, 1, 1}=\langle -b_{1,5} - b_{2,6}+ b_{3,7} + b_{4,8} \rangle\\
      L_{ 1,
      1, -1, -1, -1}=\langle -b_{1,4} + b_{2,3}- b_{5,8} + b_{6,7} \rangle&
      L_{ 1, 1, -1, -1, 1}=\langle -b_{1,7} + b_{2,8}- b_{3,5} + b_{4,6} \rangle\\
      L_{ 1, 1, -1, 1, -1}=\langle b_{1,8} + b_{2,7}+ b_{3,6} + b_{4,5} \rangle&
      L_{ 1,
      1, -1, 1, 1}=\langle b_{1,7} + b_{2,8}+ b_{3,5} + b_{4,6} \rangle\\
      L_{ 1, 1, 1, -1, -1}=\langle b_{1,2} + b_{3,4}+ b_{5,6} + b_{7,8} \rangle&
      L_{ 1, 1, 1, -1, 1}=\langle -b_{1,5} + b_{2,6}- b_{3,7} + b_{4,8} \rangle\\
      L_{ 1, 1, 1,
       1, -1}=\langle b_{1,6} + b_{2,5}+ b_{3,8} + b_{4,7} \rangle&
       L_{ 1, 1, 1, 1, 1}=\langle b_{1,5} + b_{2,6}+ b_{3,7} + b_{4,8} \rangle
\end{array}$$} {\small
    $$\begin{array}{l}L_{ \frac{1}{4}, -1, -1, -1,
1}=\langle b_{1, 3} + i b_{1, 7} -
    b_{2, 4} - i b_{2, 8} - i b_{3, 5} + i b_{4, 6} - b_{5, 7} + b_{6, 8} \rangle\\
L_{ \frac{1}{4}, -1, -1,
      1, -1}=\langle -b_{1, 4} - i b_{1, 8} -
    b_{2, 3} - i b_{2, 7} + i b_{3, 6} + i b_{4, 5} + b_{5, 8} + b_{6, 7} \rangle\\
      L_{ \frac{1}{4}, -1, -1, 1, 1}=\langle -b_{1, 3} - i b_{1, 7} -
    b_{2, 4} - i b_{2, 8} + i b_{3, 5} + i b_{4, 6} + b_{5, 7} + b_{6, 8} \rangle\\
      L_{ \frac{1}{4}, -1,
      1, -1, -1}=\langle b_{1, 2} + i b_{1, 6} -i
    b_{2, 5} -  b_{3, 4} - i b_{3, 8} + i b_{4, 7} - b_{5, 6} + b_{7,8} \rangle\\
      L_{ \frac{1}{4}, 1, -1, -1, -1}=\langle b_{1, 4} + i b_{1, 8} -
    b_{2, 3} - i b_{2, 7} + i b_{3, 6} - i b_{4, 5} - b_{5, 8} + b_{6, 7} \rangle\\
      L_{ \frac{1}{4}, 1,
      1, -1, -1}=\langle -b_{1, 2} - i b_{1, 6} +i
    b_{2, 5} -  b_{3, 4} - i b_{3, 8} + i b_{4, 7} + b_{5, 6} + b_{7,8} \rangle\\
    L_{ 4, -1, -1, -1, 1}=\langle b_{1, 3} - i b_{1, 7} -
    b_{2, 4} + i b_{2, 8} + i b_{3, 5} - i b_{4, 6} - b_{5, 7} + b_{6, 8} \rangle\\
       L_{ 4, -1, -1,
      1, -1}=\langle -b_{1, 4} + i b_{1, 8} -
    b_{2, 3}+ i b_{2, 7} - i b_{3, 6} - i b_{4, 5} + b_{5, 8} + b_{6, 7} \rangle\\
      L_{ 4, -1, -1, 1, 1}=\langle -b_{1, 3} + i b_{1, 7} -
    b_{2, 4} + i b_{2, 8} - i b_{3, 5} - i b_{4, 6} + b_{5, 7} + b_{6, 8} \rangle\\
      L_{ 4, -1, 1, -1, -1}=\langle b_{1, 2} - i b_{1, 6} +
    ib_{2, 5} -  b_{3, 4} + i b_{3, 8} - i b_{4, 7} - b_{5, 6} + b_{7,8} \rangle\\
      L_{ 4,
      1, -1, -1, -1}=\langle b_{1, 4} - i b_{1, 8} -
    b_{2, 3} + i b_{2, 7} - i b_{3, 6} + i b_{4, 5} - b_{5, 8} + b_{6, 7} \rangle\\
      L_{ 4, 1, 1, -1, -1  }=\langle -b_{1, 2} + i b_{1, 6} -
    ib_{2, 5} -  b_{3, 4}+i b_{3, 8} -i b_{4, 7} + b_{5, 6} + b_{7,8}
    \rangle\end{array}$$}

\textbf{Grading over $ \mathbb Z_2^7$:} $Q_5=\langle\{ F_i\mid
i=1,\dots,7\}\rangle\cong \mathbb Z_2^7$
  is  a MAD of $\mathop{\rm Aut} L$, corresponding to $T_{ 0,8}^{( 0)}$.
The simultaneous diagonalization of $L$ relative to it is   of
type $(28)$: {\small $$\begin{array}{llll} L_{ 1, -1, -1, 1, 1, 1,
1}=\langle b_{ 6,7 } \rangle & L_{ 1, -1, 1, -1, 1, 1, 1}=\langle
b_{5 ,7 }\rangle &L_{ 1, -1, 1, 1, -1,
       1, 1}=\langle b_{ 4,7 } \rangle
       &L_{ 1, -1, 1, 1, 1, -1, 1}=\langle  b_{ 3,7 }\rangle
       \\L_{ 1, -1, 1, 1, 1, 1, -1}=\langle b_{ 2,7 } \rangle
       &L_{
      1, -1, 1, 1, 1, 1, 1}=\langle b_{1 ,7 }  \rangle
      & L_{ 1, 1, -1, -1, 1, 1, 1}=\langle b_{5 ,6 } \rangle
      &L_{ 1, 1, -1,
      1, -1, 1, 1}=\langle b_{ 4,6 } \rangle
      \\ L_{ 1, 1, -1, 1, 1, -1, 1}=\langle b_{ 3,6 } \rangle&
      L_{ 1, 1, -1, 1, 1,
      1, -1}=\langle b_{ 2,6 } \rangle
      & L_{ 1, 1, -1, 1, 1, 1, 1}=\langle b_{ 1,6 } \rangle&
      L_{ 1, 1, 1, -1, -1, 1, 1}=\langle b_{ 4,5 } \rangle\\
      L_{ 1,
       1, 1, -1, 1, -1, 1}=\langle b_{ 3,5 } \rangle &
       L_{ 1, 1, 1, -1, 1, 1, -1}=\langle b_{ 2,5 } \rangle &
       L_{ 1, 1, 1, -1, 1,
      1, 1}=\langle b_{ 1,5 }  \rangle &
      L_{ 1, 1, 1, 1, -1, -1, 1}=\langle b_{ 3,4 } \rangle\\
      L_{ 1, 1, 1, 1, -1, 1, -1}=\langle b_{ 2,4 } \rangle&
       L_{ 1,
       1, 1, 1, -1, 1, 1}=\langle  b_{ 1,4 }\rangle&
       L_{ 1, 1, 1, 1, 1, -1, -1}=\langle  b_{ 2,3 }\rangle&
       L_{ 1, 1, 1, 1,
      1, -1, 1}=\langle b_{1 ,3 } \rangle\\
      L_{ 1, 1, 1, 1, 1, 1, -1}=\langle b_{1 ,2 } \rangle
      &L_{ -1, -1, 1, 1, 1, 1,
      1}=\langle b_{ 7,8 } \rangle &
      L_{ -1, 1, -1, 1, 1, 1, 1}=\langle  b_{ 6,8 }\rangle&
      L_{ -1, 1, 1, -1, 1, 1, 1}=\langle b_{5 ,8 } \rangle\\
      L_{ -1,
      1, 1, 1, -1, 1, 1}=\langle b_{4 ,8 }  \rangle&
      L_{ -1, 1, 1, 1, 1, -1, 1}=\langle b_{ 3, 8} \rangle
      &L_{ -1, 1, 1, 1, 1,
      1, -1}=\langle  b_{ 2,8 }\rangle&
      L_{ -1, 1, 1, 1, 1, 1, 1  }=\langle b_{ 1, 8} \rangle
\end{array}$$}

\textbf{Grading over $\mathbb Z\times \mathbb Z_2^5$:}
$Q_6=\langle\{ p(u),F_8,F_7,F_6,F_5,F_4\mid u\in\mathbb
C^*\}\rangle\cong\mathbb C^*\times \mathbb Z_2^5$
  is again an abelian diagonalizable subgroup of $\mathop{\rm Aut} L$,
  concretely the MAD  $T_{ 2,6}^{( 0)}$.
The simultaneous diagonalization of $L$ relative to the quasitorus
$Q_6$ is of type $(28 )$: {\small $$\begin{array}{ll}L_{
\frac{1}{2}, -1, 1, 1, 1, 1}=\langle i b_{1, 7} + b_{1, 8}\rangle
&L_{\frac{1}{2}, 1, -1, 1, 1,
      1}=\langle i b_{2, 7} + b_{2, 8}\rangle \\
      L_{\frac{1}{2}, 1, 1, -1, 1, 1}=\langle i b_{3, 7} + b_{3, 8}\rangle
      &L_{\frac{1}{2}, 1, 1, 1, -1,
      1}=\langle i b_{4, 7} + b_{4, 8}\rangle
      \\ L_{\frac{1}{2}, 1, 1, 1, 1, -1}=\langle i b_{5, 7} + b_{5, 8}\rangle
      &L_{\frac{1}{2}, 1, 1, 1, 1,
      1}=\langle i b_{6, 7} + b_{6, 8}\rangle
      \\
      L_{2, -1, 1, 1, 1, 1}=\langle -i b_{1, 7} + b_{1, 8}\rangle &
      L_{2, 1, -1, 1, 1, 1}=\langle -i b_{2, 7} + b_{2, 8}\rangle \\
      L_{2, 1, 1, -1,
      1, 1}=\langle -i b_{3, 7} + b_{3, 8}\rangle &
      L_{2, 1, 1, 1, -1, 1}=\langle -i b_{4, 7} + b_{4, 8}\rangle \\
      L_{2, 1, 1, 1, 1, -1}=\langle -i b_{5, 7} + b_{5, 8}\rangle &
      L_{2, 1, 1,
      1, 1, 1  }=\langle -i b_{6, 7} + b_{6, 8}
      \rangle\end{array}$$}
      {\small $$\begin{array}{llll}L_{1, -1, -1, 1, 1, 1}=\langle b_{1, 2}\rangle
      &L_{1, -1, 1, -1, 1, 1}=\langle b_{1, 3}\rangle
      & L_{1, -1, 1,
      1, -1, 1}=\langle b_{1, 4}\rangle
      &L_{1, -1, 1, 1, 1, -1}=\langle b_{1, 5}\rangle\\
      L_{1, -1, 1, 1, 1, 1}=\langle b_{1, 6}\rangle &
      L_{1,
      1, -1, -1, 1, 1}=\langle b_{2, 3}\rangle &
      L_{1, 1, -1, 1, -1, 1}=\langle b_{2, 4}\rangle &
      L_{1, 1, -1, 1,
      1, -1}=\langle b_{2, 5}\rangle \\
      L_{1, 1, -1, 1, 1, 1}=\langle b_{2, 6}\rangle &
      L_{1, 1, 1, -1, -1, 1}=\langle b_{3, 4}\rangle & L_{1, 1,
      1, -1, 1, -1}=\langle b_{3, 5}\rangle &
      L_{1, 1, 1, -1, 1, 1}=\langle b_{3, 6}\rangle \\
      L_{1, 1, 1, 1, -1, -1}=\langle b_{  4,5}\rangle &
      L_{      1, 1, 1, 1, -1, 1}=\langle b_{  4,6}\rangle &
      L_{1, 1, 1, 1, 1, -1}=\langle b_{  5,6}\rangle &L_{1, 1, 1, 1, 1,
      1}=\langle b_{  7,8}\rangle\end{array}$$}

\textbf{Grading over $\mathbb Z^2\times \mathbb Z_2^3$:}
$Q_7=\langle\{ p(u),F_8,F_7,F_6,q(v)\mid u,v\in\mathbb
C^*\}\rangle\cong(\mathbb C^*)^2\times \mathbb Z_2^3$
  is  a MAD of $\mathop{\rm Aut} L$, corresponding to  $T_{ 4,4}^{( 0)}$,
  producing the following grading,
   of type $(26,1 )$:
{\small $$\begin{array}{ll}  L_{ 2, 1, 1, 1, 3}=\langle -b_{ 5,7
}- ib_{5 ,8 }-ib_{ 6,7 }+ b_{6 ,8 }\rangle& L_{ \frac{1}{2}, 1, 1,
      1, \frac{1}{3}}=
      \langle -b_{ 5,7 }+ ib_{5 ,8 }+ib_{ 6,7 }+ b_{6 ,8 }
      \rangle\\L_{ 2, 1, 1, 1, \frac{1}{3}}=\langle  b_{ 5,7 }+ ib_{5 ,8 }-ib_{
6,7 }+ b_{6 ,8 } \rangle& L_{ \frac{1}{2}, 1, 1,
      1, 3}=\langle  b_{ 5,7 }- ib_{5 ,8 }+ib_{ 6,7 }+ b_{6 ,8 }
      \rangle\end{array} $$
      $$\begin{array}{lll} L_{
\frac{1}{2}, -1, 1, 1, 1}=\langle   ib_{ 1,7 }+ b_{1 ,8 } \rangle&
L_{ \frac{1}{2}, 1, -1, 1,
      1}=\langle ib_{ 2,7 }+ b_{2 ,8 } \rangle&
      L_{ \frac{1}{2}, 1, 1, -1, 1}=\langle ib_{ 3,7 }+ b_{3 ,8 }
      \rangle\\
      L_{ \frac{1}{2}, 1, 1, 1, 1}=\langle ib_{ 4,7 }+ b_{4 ,8 } \rangle&
L_{ 1, -1, -1, 1, 1}=\langle b_{1 ,2 } \rangle&
      L_{ 1, -1, 1, -1, 1}=\langle b_{1 ,3 } \rangle\\
      L_{ 1, -1, 1,
      1, \frac{1}{3}}=\langle ib_{ 1,5 }+ b_{1 ,6 } \rangle&
      L_{ 1, -1, 1, 1, 1}=\langle b_{1 ,4 } \rangle&
      L_{ 1, -1, 1, 1, 3}=\langle -ib_{ 1,5 }+ b_{1 ,6 }  \rangle\\ L_{ 1,
      1, -1, -1, 1}=\langle b_{2,3 } \rangle&
      L_{ 1, 1, -1, 1, \frac{1}{3}}=\langle ib_{ 2,5 }+ b_{2 ,6 }  \rangle& L
      _{ 1, 1, -1, 1,
      1}=\langle b_{2,4 } \rangle\\
      L_{ 1, 1, -1, 1, 3}=\langle -ib_{ 2,5 }+ b_{2 ,6 }  \rangle&
      L_{ 1, 1, 1, -1, \frac{1}{3}}=\langle ib_{ 3,5 }+ b_{3 ,6 }
      \rangle&
      L_{ 1, 1,
      1, -1, 1}=\langle b_{3,4 } \rangle\\
      L_{ 1, 1, 1, -1, 3}=\langle -ib_{ 3,5 }+ b_{3 ,6 }  \rangle&
      L_{ 1, 1, 1, 1, \frac{1}{3}}=\langle ib_{ 4,5 }+ b_{4 ,6 }  \rangle& L_{ 1,
      1, 1, 1, 1}=\langle b_{5,6 },\,b_{7,8} \rangle\\
      L_{ 1, 1, 1, 1, 3}=\langle -ib_{ 4,5 }+ b_{4 ,6 }  \rangle&
      L_{ 2, -1, 1, 1, 1}=\langle -ib_{ 1,7 }+ b_{1 ,8 } \rangle& L_{ 2, 1, -1,
      1, 1}=\langle -ib_{ 2,7 }+ b_{2 ,8 } \rangle\\
      L_{ 2, 1, 1, -1, 1}=\langle -ib_{ 3,7 }+ b_{3 ,8 } \rangle&
       L_{ 2, 1,
      1, 1, 1}=\langle -ib_{ 4,7 }+ b_{4 ,8 } \rangle&
\end{array}$$}

\textbf{Grading over $\mathbb Z_2^3\times\mathbb Z_4$:}
$Q_{8}=\langle\{G_8,G_{10},G_{11},G_{12}\}\rangle\cong\mathbb
Z_2^3\times\mathbb Z_4$
  is  a MAD of $\mathop{\rm Aut} L$, corresponding to $T_{ 0,2}^{( 2)}$, which
  produces in
   $L$ a grading  of type $(24,2 )$:{\small $$\begin{array}{l} L_{ -1,
-1, -1, - i  }=\langle i b_{1, 7}+ b_{1, 8}+i b_{2, 7} -
    b_{2, 8} - i b_{3, 5} -
    b_{3, 6} - i b_{4, 5} + b_{4, 6}\rangle\\L_{
-1, -1, -1,  i  }=\langle -i b_{1, 7}+ b_{1, 8}-i b_{2, 7} -
    b_{2, 8} + i b_{3, 5} -
    b_{3, 6} + i b_{4, 5} + b_{4, 6}\rangle\\L_{ -1, -1, 1, - i }=\langle -i b_{1, 7}- b_{1, 8}+i
b_{2, 7} -
    b_{2, 8} + i b_{3, 5} +
    b_{3, 6} - i b_{4, 5} + b_{4, 6}\rangle\\L_{ -1, -1,
       1,  i  }=\langle i b_{1,
7}- b_{1, 8}-i b_{2, 7} -
    b_{2, 8} - i b_{3, 5} +
    b_{3, 6} + i b_{4, 5} + b_{4, 6}\rangle\\
    L_{ -1,
      1, -1, - i  }=\langle i b_{1,
5}+ b_{1, 6}+i b_{2, 5} -
    b_{2, 6} - i b_{3, 7} -
    b_{3, 8} - i b_{4, 7} + b_{4, 8}\rangle\\L_{ -1, 1, -1,  i  }=\langle -i b_{1,
5}+ b_{1, 6}-i b_{2, 5} -
    b_{2, 6} + i b_{3, 7} -
    b_{3, 8} + i b_{4, 7} + b_{4, 8}\rangle\\
    L_{ -1,
      1, 1, - i  }=\langle -i b_{1,
5}- b_{1, 6}+i b_{2, 5} -
    b_{2, 6} + i b_{3, 7} +
    b_{3, 8} - i b_{4, 7} + b_{4, 8}\rangle\\L_{ -1, 1, 1,  i  }=\langle i b_{1,
5}- b_{1, 6}-i b_{2, 5} -
    b_{2, 6} - i b_{3, 7} +
    b_{3, 8} + i b_{4, 7} + b_{4, 8}\rangle\\
    L_{
      1, -1, -1, - i  }=\langle -i b_{1,
7}- b_{1, 8}-i b_{2, 7} +
    b_{2, 8} - i b_{3, 5} -
    b_{3, 6} - i b_{4, 5} + b_{4, 6}\rangle\\L_{ 1, -1, -1,  i  }=\langle i b_{1,
7}- b_{1, 8}+i b_{2, 7} +
    b_{2, 8} + i b_{3, 5} -
    b_{3, 6} + i b_{4, 5} + b_{4, 6}\rangle\\
    L_{
      1, -1, 1, - i  }=\langle i b_{1,
7}+ b_{1, 8}-i b_{2, 7} +
    b_{2, 8} + i b_{3, 5} +
    b_{3, 6} - i b_{4, 5} + b_{4, 6}\rangle\\L_{ 1, -1, 1,  i  }=\langle -i b_{1,
7}+ b_{1, 8}+i b_{2, 7} +
    b_{2, 8} - i b_{3, 5} +
    b_{3, 6} + i b_{4, 5} + b_{4, 6}\rangle\\
    L_{ 1,
      1, -1, - i  }=\langle -i b_{1,
5}- b_{1, 6}-i b_{2, 5} +
    b_{2, 6} - i b_{3, 7} -
    b_{3, 8} - i b_{4, 7} + b_{4, 8}\rangle\\L_{ 1, 1, -1,  i  }=\langle i b_{1,
5}- b_{1, 6}+i b_{2, 5} +
    b_{2, 6} + i b_{3, 7} -
    b_{3, 8} + i b_{4, 7} + b_{4, 8}\rangle\\
   L_{ 1, 1,
      1, - i  }=\langle i b_{1,
5}+ b_{1, 6}-i b_{2, 5} +
    b_{2, 6} + i b_{3, 7} +
    b_{3, 8} - i b_{4, 7} + b_{4, 8}\rangle\\L_{ 1, 1, 1,  i    }=\langle -i b_{1,
5}+ b_{1, 6}+i b_{2, 5} +
    b_{2, 6} - i b_{3, 7} +
    b_{3, 8} + i b_{4, 7} + b_{4, 8}\rangle
\end{array}$$
 $$\begin{array}{ll} L_{ -1, -1, -1, -1}=\langle b_{5 ,  8}
+ b_{6 , 7 }\rangle& L_{ -1, -1, -1, 1}=
    \langle -b_{ 1,3  } + b_{2 , 4 }\rangle\\
    L_{ -1, 1, -1, -1}=\langle -b_{ 1,2  } + b_{3 ,4  }\rangle&
    L_{ -1, 1, -1, 1}=\langle -b_{ 5, 6 } + b_{7 , 8 }\rangle\\
    L_{ 1, -1, -1, 1}=\langle -b_{ 5, 8 } + b_{6 , 7 }\rangle&
     L_{ 1, 1, -1, 1}=\langle b_{ 5, 6 } + b_{ 7, 8 }\rangle\\
     L_{ 1, 1, -1, -1}=\langle b_{ 1,2  } + b_{3 , 4 }\rangle&
     L_{ -1, -1, 1,
-1}=\langle b_{1 , 4 } + b_{ 2, 3 } ,\,- b_{ 5, 7
}+b_{6 , 8} \rangle\\
L_{ 1, -1, -1, -1}=\langle -b_{1 , 4 } + b_{ 2, 3 }\rangle& L_{
-1, -1, 1, 1}=\langle b_{1 , 3 } + b_{ 2, 4 } ,\,b_{5 , 7 } + b_{
6, 8 }\rangle
\end{array}$$}

\textbf{Grading over $\mathbb Z^3\times \mathbb Z_2$:}
$Q_9=\langle\{ p(u),F_8,r(v),q(w)\mid u,v,w\in\mathbb
C^*\}\rangle\cong(\mathbb C^*)^3\times \mathbb Z_2$
  is an abelian diagonalizable subgroup of $\mathop{\rm Aut} L$,
  concretely is the MAD-group $T_{ 6,2}^{( 0)}$, which
  produces a grading  of type $(25,0,1 )$:

{\small $$\begin{array}{ll}  L_{\frac{1}{2}, 1, 1, \frac{1}{3}}=
      \langle -b_{ 5,7 }+ ib_{5 ,8 }+ib_{ 6,7 }+ b_{6 ,8 } \rangle &L_{\frac{1}{2}, 1, \frac{1}{5},
      1}=\langle -b_{ 3,7 }+ ib_{3 ,8 }+ib_{ 4,7 }+ b_{4 ,8 } \rangle \\
L_{\frac{1}{2}, 1, 1, 3}=\langle b_{ 5,7 }-ib_{5 ,8 }+ib_{ 6,7 }+
b_{6 ,8 } \rangle & L
      _{\frac{1}{2}, 1, 5, 1}=\langle b_{ 3,7 }- ib_{3 ,8 }+ib_{ 4,7 }+ b_{4 ,8 } \rangle \\
      L_{1,
      1, \frac{1}{5}, \frac{1}{3}}=\langle -b_{ 3,5 }+ ib_{3 ,6 }+ib_{ 4,5 }+ b_{4 ,6 }  \rangle
      &
L_{1, 1, 5, \frac{1}{3}}=\langle  b_{ 3,5 }- ib_{3 ,6 }+ib_{ 4,5 }+ b_{4 ,6 }  \rangle \\
L_{1,
      1, \frac{1}{5}, 3}=\langle  b_{ 3,5 }+ ib_{3 ,6 }-ib_{ 4,5 }+ b_{4 ,6 }  \rangle
      &
L_{1, 1, 5,
       3}=\langle -b_{ 3,5 }- ib_{3 ,6 }-ib_{ 4,5 }+ b_{4 ,6 } \rangle \\
       L_{2, 1, \frac{1}{5}, 1}=\langle  b_{ 3,7 }+ ib_{3 ,8 }-ib_{ 4,7 }+ b_{4 ,8 } \rangle
       &
       L_{2, 1,
      1, \frac{1}{3}}=\langle  b_{ 5,7 }+ ib_{5 ,8 }-ib_{ 6,7 }+ b_{6 ,8 } \rangle
      \\
      L_{2, 1, 1, 3}=\langle -b_{ 5,7 }- ib_{5 ,8 }-ib_{ 6,7 }+ b_{6 ,8 } \rangle &
      L_{2, 1, 5, 1}=\langle -b_{ 3,7 }- ib_{3 ,8 }-ib_{ 4,7 }+ b_{4 ,8 } \rangle\end{array}$$ $$\begin{array}{lll} L_{
\frac{1}{2}, -1, 1, 1}=\langle ib_{ 1,7 }+ b_{1 ,8 } \rangle &
L_{\frac{1}{2}, 1, 1,1}=\langle ib_{ 2,7 }+ b_{2 ,8 } \rangle &
L_{      1, -1, \frac{1}{5}, 1}=\langle ib_{ 1,3 }+ b_{1 ,4 }
\rangle \\
      L_{1, -1, 1, \frac{1}{3}}=\langle ib_{ 1,5 }+ b_{1 ,6 } \rangle
      &
      L_{1, -1, 1,
      1}=\langle b_{1,2} \rangle &
      L_{1, -1, 1, 3}=\langle -ib_{ 1,5 }+ b_{1 ,6 } \rangle \\
      L_{1, -1, 5, 1}=\langle -ib_{ 1,3 }+ b_{1 ,4 } \rangle &
L_{1, 1, \frac{1}{5}, 1}=\langle ib_{ 2,3 }+ b_{2 ,4 } \rangle &
L_{1, 1, 1, \frac{1}{3}}=\langle ib_{ 2,5 }+ b_{2 ,6 } \rangle \\
      L_{1, 1, 1, 1}=\langle b_{3,4},\,b_{5,6},\,b_{7,8} \rangle &
      L_{      1, 1, 1, 3}=\langle -ib_{ 2,5 }+ b_{2 ,6 } \rangle &
L_{1, 1, 5, 1}=\langle -ib_{ 2,3 }+ b_{2 ,4 } \rangle \\
L_{2, -1, 1, 1}=\langle -ib_{ 1,7 }+ b_{1 ,8 }\rangle & L_{2, 1,
1, 1}=\langle -ib_{ 2,7 }+ b_{2 ,8 } \rangle &
\end{array}$$}

\textbf{Grading over $\mathbb Z^4$:} $Q_{10}=\langle\{\mathop{\rm
Ad} p(u),\mathop{\rm Ad} s(v),\mathop{\rm Ad} r(w),\mathop{\rm Ad}
q(z)\mid u,v,w,z\in\mathbb C^*\}\rangle\cong(\mathbb C^*)^4$
  is an abelian diagonalizable subgroup of $\mathop{\rm Aut} L$ (corresponding to $T_{ 8,0}^{( 0)}$).
The simultaneous diagonalization of $L$ relative to the quasitorus
$Q_{10}$ is of type $(24,0,0,1 )$. Of course it is the
\emph{Cartan grading}, in other words, the decomposition in root
spaces relative to the Cartan subalgebra $L_{1,1,1,1}=\langle b_{
1,2 },\,b_{3 ,4 },\,b_{ 5,6 },\,b_{ 7,8 }\rangle$: {\small
$$\begin{array}{ll} L_{ \frac12,\frac17,1,1}=\langle -b_{ 1,7 }+i
b_{1 ,8 }+ib_{ 2,7 }+b_{2 , 8}\rangle& L_{
\frac12,1,\frac15,1}=\langle -b_{ 3,7 }+i b_{3,8 }+ib_{ 4,7 }+b_{4
, 8}
\rangle \\
L_{\frac12,1,1,\frac13 }=\langle -b_{ 5,7 }+i b_{5 ,8 }+ib_{ 6,7
}+b_{6 ,8 }\rangle&L_{\frac12,1,1,3 }=\langle b_{ 5,7 }-i b_{5 ,8
}+ib_{ 6,7 }+b_{6 ,8 }\rangle\\L_{ \frac12,1, 5,1}=\langle b_{ 3,7
}-i b_{3,8 }+ib_{ 4,7 }+b_{4 , 8}\rangle &L_{
\frac12,7,1,1}=\langle b_{ 1,7 }-i b_{1 ,8 }+ib_{ 2,7 }+b_{2 ,
8}\rangle\\
L_{ 1,\frac17,\frac15,1}=\langle -b_{ 1,3 }+i b_{1 ,4 }+ib_{ 2,3
}+b_{2 , 4}\rangle& L_{ 1,\frac17,1,\frac13}=\langle -b_{1,5 }+i
b_{1,6 }+ib_{ 2,5 }+b_{2 , 6}\rangle\\
L_{ 1,\frac17, 5,1}=\langle b_{ 1,3 }+i b_{1 ,4 }-ib_{ 2,3 }+b_{2
, 4}\rangle& L_{ 1,\frac17,1,3}=\langle b_{1,5 }+i b_{1,6 }-ib_{
2,5 }+b_{2 , 6}\rangle\\
L_{1, 1,\frac15,\frac13}=\langle -b_{ 3,5 }+i b_{3,6 }+ib_{4,5
}+b_{4,6}\rangle& L_{ 1,1,\frac15,3}=\langle b_{ 3,5 }+i b_{3,6
}-ib_{4,5 }+b_{4,6}\rangle\\
L_{1, 1, 5,\frac13}=\langle b_{ 3,5 }-i b_{3,6 }+ib_{4,5
}+b_{4,6}\rangle& L_{ 1,1, 5,3}=\langle -b_{ 3,5 }-i b_{3,6
}-ib_{4,5 }+b_{4,6}\rangle\\
L_{ 1, 7,\frac15,1}=\langle  b_{ 1,3 }-i b_{1 ,4 }+ib_{ 2,3 }+b_{2
, 4}\rangle& L_{ 1, 7,1,\frac13}=\langle  b_{1,5 }-i b_{1,6 }+ib_{
2,5 }+b_{2 , 6}\rangle\\
L_{ 1, 7, 5,1}=\langle -b_{ 1,3 }-i b_{1 ,4 }-ib_{ 2,3 }+b_{2 ,
4}\rangle& L_{ 1, 7,1, 3}=\langle -b_{1,5 }-i b_{1,6 }-ib_{ 2,5
}+b_{2 , 6}\rangle\\
L_{  2,\frac17,1,1}=\langle  b_{ 1,7 }+i b_{1 ,8 }-ib_{ 2,7 }+b_{2
, 8}\rangle& L_{  2,1,\frac15,1}=\langle  b_{ 3,7 }+i b_{3,8
}-ib_{ 4,7 }+b_{4 , 8}\rangle\\
L_{ 2,1,1,\frac13 }=\langle  b_{ 5,7 }+i b_{5 ,8 }-ib_{ 6,7 }+b_{6
,8 }\rangle&L_{ 2,1,1,3 }=\langle -b_{ 5,7 }-i b_{5 ,8 }-ib_{ 6,7
}+b_{6 ,8 }\rangle\\L_{  2,1, 5,1}=\langle -b_{ 3,7 }-i b_{3,8
}-ib_{ 4,7 }+b_{4 , 8}\rangle &L_{  2,7,1,1}=\langle -b_{ 1,7 }-i
b_{1 ,8 }-ib_{ 2,7 }+b_{2 , 8}\rangle\end{array}$$}

\textbf{Grading over $\mathbb Z_2^6$:}
$Q_{11}=\langle\{G_8,G_{10},G_{13},G_{3},G_{7},G_{14}
\}\rangle\cong\mathbb Z_2^6$
  is also a MAD of $\mathop{\rm Aut} L$, corresponding to $T_{ 0,1}^{( 3)}$.
The induced grading is of type $(28 )$:

{\small $$\begin{array}{ll} L_{ -1, -1, -1, -1, -1, -1}=\langle
b_{1, 8} - b_{2, 7} - b_{3, 6} + b_{4, 5}\rangle & L_{ -1, -1, -1,
-1, 1, 1}=\langle b_{1, 7} - b_{2, 8} - b_{3, 5} + b_{4, 6}\rangle
\\ L_{ -1, -1, -1,
      1, -1, 1}=\langle -b_{1, 8} - b_{2, 7} + b_{3,
6} + b_{4, 5}\rangle&
       L_{ -1, -1, -1, 1, 1, 1}=\langle -b_{1, 7}
- b_{2, 8} + b_{3, 5} + b_{4, 6}\rangle \\
        L_{ -1, -1, 1, -1,
      1, -1}=\langle b_{1, 3}
- b_{2, 4} - b_{5,7} + b_{6,8}\rangle &
       L_{ -1, -1, 1, -1, 1, 1}=\langle -b_{1, 3}
+ b_{2, 4} - b_{5,7} + b_{6,8}\rangle \\
        L_{ -1, -1, 1,
      1, -1, -1}=\langle -b_{1, 4}
- b_{2, 3} + b_{5,8} + b_{6,7}\rangle &
       L_{ -1, -1, 1, 1, -1, 1}=\langle b_{1, 4}
+ b_{2, 3} + b_{5,8} + b_{6,7}\rangle \\
        L_{ -1, -1, 1, 1,
      1, -1}=\langle -b_{1, 3}
- b_{2, 4} + b_{5,7} + b_{6,8}\rangle &
       L_{ -1, -1, 1, 1, 1, 1}=\langle b_{1, 3}
+ b_{2, 4} + b_{5,7} + b_{6,8}\rangle \\
        L_{ -1, 1, -1, -1, -1, 1}=\langle b_{1, 6}
- b_{2, 5} - b_{3,8} + b_{4,7}\rangle &
         L_{ -1,
      1, -1, -1, 1, -1}=\langle b_{1, 5}
- b_{2, 6} - b_{3,7} + b_{4,8}\rangle \\
       L_{ -1, 1, -1, 1, -1, -1}=\langle -b_{1, 6}
- b_{2, 5} + b_{3,8} + b_{4,7}\rangle &
        L_{ -1, 1, -1, 1,
      1, -1}=\langle -b_{1, 5}
- b_{2, 6} + b_{3,7} + b_{4,8}\rangle \\
       L_{ -1, 1, 1, -1, -1, -1}=\langle b_{1, 2}
- b_{3,4} - b_{5,6} + b_{7,8}\rangle &
        L_{ -1, 1, 1, -1, -1, 1}=\langle -b_{1, 2}
+ b_{3,4} - b_{5,6} + b_{7,8}\rangle \\
         L_{
      1, -1, -1, -1, -1, 1}=\langle -b_{1, 8} + b_{2, 7} - b_{3,
6} + b_{4, 5}\rangle &
       L_{ 1, -1, -1, -1, 1, -1}=\langle -b_{1, 7}
+ b_{2, 8} - b_{3, 5} + b_{4, 6}\rangle \\
        L_{ 1, -1, -1,
      1, -1, -1}=\langle b_{1, 8} + b_{2, 7} + b_{3,
6} + b_{4, 5}\rangle &
      L_{ 1, -1, -1, 1, 1, -1}=\langle b_{1, 7}
+ b_{2, 8} + b_{3, 5} + b_{4, 6}\rangle \\
       L_{ 1, -1, 1, -1, -1, -1}=\langle b_{1, 4}
- b_{2, 3} - b_{5,8} + b_{6,7}\rangle &
        L_{
      1, -1, 1, -1, -1, 1}=\langle -b_{1, 4}
+ b_{2, 3} - b_{5,8} + b_{6,7}\rangle \\
       L_{ 1, 1, -1, -1, -1, 1}=\langle -b_{1, 6}
+ b_{2, 5} - b_{3,8} + b_{4,7}\rangle &
        L_{ 1, 1, -1, -1,
      1, -1}=\langle -b_{1, 5}
+ b_{2, 6} - b_{3,7} + b_{4,8}\rangle \\
       L_{ 1, 1, -1, 1, -1, -1}=\langle b_{1, 6}
+ b_{2, 5} + b_{3,8} + b_{4,7}\rangle &
        L_{ 1, 1, -1, 1, 1, -1}=\langle b_{1, 5}
+ b_{2, 6} + b_{3,7} + b_{4,8}\rangle \\
         L_{ 1, 1,
      1, -1, -1, -1}=\langle -b_{1, 2}
- b_{3,4} + b_{5,6} + b_{7,8}\rangle &
       L_{ 1, 1, 1, -1, -1, 1}=\langle b_{1, 2}
+ b_{3,4} + b_{5,6} + b_{7,8}\rangle
\end{array}$$}

 In order to describe the gradings involving a copy of $\mathbb Z_3$, we introduce
new automorphisms. Fix $\mathcal{B}=\{b_1,\dots,b_{28}\}$ a basis
of $L$ of root vectors (take the basis produced by $Q_{10}$,
suitably ordered, to express the remaining three gradings also in
terms of $b_{ij}$), and let $\varphi_{i,j}(b_k)=\delta_{k,j}b_i$
for $i,j,k=1\dots28$. Define {\small
\begin{align*} H_1 = & \,\varphi_{2,4}+ \varphi_{3,1}+
\varphi_{3,2}+ \varphi_{5,27}+ \varphi_{6,8}+ \varphi_{7,9}+
\varphi_{ 8,25} +
  \varphi_{11,12}+\varphi_{12,26}+\varphi_{ 13,18}+\varphi_{14,23} \\ &+\varphi_{17,15}+\varphi_{18,20}+
  \varphi_{ 19,21}+\varphi_{ 20,13}+\varphi_{ 23,24} +\varphi_{ 24,14}+\varphi_{ 25,6}+\varphi_{ 26,11}-\varphi_{1,1}
-\varphi_{1,2} -\varphi_{1,3} \\ &-\varphi_{1,4} -\varphi_{4,2}
-\varphi_{4,4} -\varphi_{9,22} -\varphi_{10,19} -\varphi_{15,28}
-\varphi_{16,5} -\varphi_{21,10} -\varphi_{22,7} -\varphi_{27,16}
-\varphi_{28,17}\\  H_2= & \,\varphi_{2,2}+\varphi_{
2,3}+\varphi_{ 2,4}+\varphi_{ 4,1}+\varphi_{ 5,26}+\varphi_{
6,16}+\varphi_{7,19}+\varphi_{ 8,5}+\varphi_{ 10,22}+\varphi_{
11,13}+\varphi_{ 12,18}+\varphi_{ 14,8}\\
&+\varphi_{15,23}+\varphi_{ 16,12}+\varphi_{ 17,14}+\varphi_{
18,28}+\varphi_{ 19,7}+\varphi_{ 20,17}+\varphi_{22,10}+\varphi_{
23,25}+\varphi_{ 24,6}+\varphi_{ 26,20}\\ &+\varphi_{
27,11}+\varphi_{ 28,24}-\varphi_{ 1,1}-\varphi_{ 1,3}-\varphi_{
1,4}-\varphi_{ 3,3}-\varphi_{ 9,21}-\varphi_{ 13,15}-\varphi_{
21,9}-\varphi_{ 25,27}-2\varphi_{1,2}\\ t_{x,y,z,u} & :=
\text{diag} \{ x,y,z,u,xy,xyz,yz,uxy,uy,uxy^2z,uxyz,uyz,{\frac1x},
{\frac1y},{\frac1z},{\frac1u},{\frac1{xy}},  {\frac1{xyz}},
\frac1{yz},\frac1{uxy},\\
&\qquad\qquad\frac1{uy},\frac1{uxy^2z},\frac1{uxyz},\frac1{uyz}\}
\end{align*}}

\textbf{Grading over $\mathbb Z_3^3$:} If we denote
$\omega=-(-1)^{\frac13}$,
$Q_{12}=\langle\{t_{1,\omega^2,\omega^2,\omega^2},t_{\omega^2,1,\omega^2,1},H_1
\}\rangle\cong\mathbb Z_3^3$
  is an abelian diagonalizable subgroup of $\mathop{\rm Aut} L$ inducing a grading of type $(24,2 )$: {\small $$\begin{array}{l}L_{ 1, 1,
\omega }=\langle -\omega b_{1}-\omega
b_{2}+b_{3},\,(1+\omega)b_{2}+b_{4} \rangle\\L_{1,
      1, {\omega^2}}=\langle -\omega^2   b_{1}-\omega^2   b_{2}+b_{3},\,(1+ \omega^2  )b_{2}+b_{4}
      \rangle\end{array}$$
$$\begin{array}{ll}
      L_{      1, \omega , 1}=\langle b_{13} + b_{18 }+ b_{20}\rangle&
      L_{      1, \omega , \omega}=\langle \omega^2  b_{13}+ \omega b_{18}+ b_{20} \rangle\\
L_{      1, \omega , \omega^2  }=\langle \omega b_{13} + \omega^2
b_{18 }+ b_{20 } \rangle&
L_{1,  \omega^2  , 1}=\langle b_{6} + b_{8}+ b_{25} \rangle\\
L_{      1,  \omega^2  , \omega}=\langle \omega^2  b_{6}+ \omega
b_{8 }+ b_{25 } \rangle&L_{
      1,  \omega^2  , \omega^2  }=\langle \omega b_{ 6} +\omega^2   b_{8 }+ b_{25 } \rangle\\
      L_{\omega , 1,
      1}=\langle -b_{ 15} - b_{17 }+ b_{28 } \rangle&L_{\omega ,
      1, \omega }=\langle  -\omega b_{ 15}- \omega^2   b_{17 }+ b_{28 }  \rangle\\L_{\omega,
      1,  \omega^2  }=\langle - \omega^2  b_{ 15}   -\omega b_{17 }+ b_{28 }  \rangle&
      L_{\omega, \omega ,
      1}=\langle -b_{ 10} + b_{19 }+ b_{21 }  \rangle\\
      L_{\omega , \omega, \omega }=\langle - \omega^2  b_{ 10}+ \omega b_{19 }+ b_{21 } \rangle&L_{\omega , \omega ,  \omega^2  }=\langle  -\omega b_{ 10} + \omega^2   b_{19 }+
b_{21 } \rangle\\L_{\omega , \omega^2  ,
      1}=\langle b_{ 14} + b_{23 }+ b_{24 }  \rangle&
      L_{\omega , \omega^2  , \omega }=\langle  \omega^2  b_{ 14} +\omega b_{23 }+ b_{24 }  \rangle\\L_{\omega ,
 \omega^2  , \omega^2  }=\langle \omega b_{ 14} + \omega^2   b_{23 }+ b_{24 } \rangle&L_{ \omega^2  , 1,
      1}=\langle b_{ 5} - b_{16 }+ b_{27 } \rangle\\L_{ \omega^2  ,
      1, \omega }=\langle \omega b_{ 5} - \omega^2   b_{16 }+ b_{27 } \rangle&
      L_{\omega^2  ,
      1,  \omega^2  }=\langle  \omega^2  b_{ 5}   -\omega b_{16 }+ b_{27 } \rangle\\
      L_{\omega^2  , \omega ,
      1}=\langle b_{ 11} + b_{12 }+ b_{26 } \rangle&
      L_{ \omega^2  , \omega, \omega }=\langle  \omega^2  b_{ 11} +\omega b_{12 }+ b_{26 } \rangle\\L_{ \omega^2  , \omega ,  \omega^2  }=\langle \omega b_{ 11} + \omega^2   b_{12 }+
b_{26 } \rangle&L_{ \omega^2  ,  \omega^2  ,
      1}=\langle  -b_{ 7} - b_{9 }+ b_{22 } \rangle\\
      L_{ \omega^2  , \omega^2  , \omega }=\langle - \omega^2  b_{ 7}   -\omega b_{9 }+ b_{22 } \rangle&L_{ \omega^2  ,
 \omega^2  , \omega^2  }=\langle   -\omega b_{ 7} - \omega^2   b_{9 }+ b_{22 } \rangle
\end{array}$$}

\textbf{Grading over $\mathbb Z^2\times\mathbb Z_3$:}
$Q_{13}=\langle\{t_{u,1,u,\frac1u},t_{1,v,\frac{1}{v^2},1},H_2^2\mid
u,v\in\mathbb C^* \}\rangle\cong(\mathbb C^*)^2\times\mathbb Z_3$
  is an abelian diagonalizable subgroup of $\mathop{\rm Aut} L$, which produces a grading of type $(26,1 )$:

{\small $$\begin{array}{ll} L_{\frac{1}{2},
      1, 1}=\langle b_{ 8}+b_{ 17}+b_{ 26} \rangle &L_{\frac{1}{2},
      1, \omega}=\langle \omega b_{ 8}+\omega^2b_{ 17}+b_{ 26} \rangle \\
      L_{\frac{1}{2},
      1, \omega^2}=\langle \omega^2b_{ 8}+\omega b_{ 17}+b_{ 26} \rangle &
      L_{\frac{1}{2}, 3,
      1}=\langle b_{ 13}-b_{ 23}+b_{ 27} \rangle \\L_{\frac{1}{2},
      3, \omega}=\langle \omega^2b_{ 13}-\omega b_{ 23}+b_{ 27} \rangle &
      L_{\frac{1}{2},
      3, \omega^2}=\langle \omega b_{ 13}-\omega^2b_{ 23}+b_{ 27} \rangle \\
      L_{      1, \frac{1}{3}, 1}=\langle b_{ 16}+b_{ 18}+b_{ 24} \rangle &
      L_{      1, \frac{1}{3}, \omega}=\langle \omega^2b_{ 16}+\omega b_{ 18}+b_{ 24}  \rangle \\
L_{1, 1, \omega}=\langle \frac{ 1+\omega^2}{1-\omega}b_{1}
+\frac{2}{1-\omega}b_{2}+\frac{1}{1-\omega}b_{3}+ b_{4}\rangle &
      L_{1,
      1, \omega^2}=\langle \frac{ 1}{1-\omega}b_{1} +\frac{2}{ 1-\omega^2}b_{2}+\frac{1}{ 1-\omega^2}b_{3}+ b_{4} \rangle \\
      L_{1, 3, 1}=\langle b_{  6}+b_{ 12}+b_{ 28}  \rangle &
      L_{1,
      3, \omega}=\langle \omega^2b_{  6}+\omega b_{ 12}+b_{ 28} \rangle \\L_{1,
      3, \omega^2}=\langle \omega b_{  6}+\omega^2 b_{ 12}+b_{ 28} \rangle &
      L_{      2, \frac{1}{3}, 1}=\langle -b_{  11}+b_{ 15}+b_{ 25} \rangle \\
      L_{      2, \frac{1}{3}, \omega}=\langle -\omega^2b_{  11}+\omega b_{ 15}+b_{ 25} \rangle &
      L_{      2, \frac{1}{3}, \omega^2}=\langle -\omega b_{  11}+\omega^2 b_{ 15}+b_{ 25} \rangle \\L_{2, 1,
      1}=\langle b_{  5}+b_{ 14}+b_{ 20} \rangle &
      L_{2, 1, \omega}=\langle \omega b_{  5}+\omega^2b_{ 14}+b_{ 20} \rangle \\
      L_{      1, \frac{1}{3}, \omega^2}=\langle \omega b_{ 16}+\omega^2b_{ 18}+b_{ 24} \rangle &
      L_{2,
      1, \omega^2}=\langle \omega^2b_{  5}+\omega b_{ 14}+b_{ 20} \rangle
      \end{array}$$
      $$\begin{array}{llll}
L_{1, 1,
      1}=\langle b_{  1}+b_{ 2},\,b_{3} \rangle &
L_{\frac{1}{4}, 3, 1}=\langle b_{22 } \rangle &\L_{\frac{1}{2},
\frac{1}{3}, 1}=\langle b_{21 } \rangle&\\
L_{\frac{1}{2}, 9, 1}=\langle b_{19} \rangle & L_{2, \frac{1}{9},
1}=\langle b_7 \rangle & L_{2, 3, 1}=\langle b_9 \rangle &
      L_{      4, \frac{1}{3}, 1  }=\langle b_{10} \rangle
\end{array}$$}

\textbf{Grading over $\mathbb Z_2^3\times\mathbb Z_3$:}
$Q_{14}=\langle\{
t_{1,-1,1,1},t_{-1,-1,-1,-1},H_2\}\rangle\cong\mathbb
Z_2^2\times\mathbb Z_6$
  is an abelian diagonalizable subgroup of $\mathop{\rm Aut} L$, producing a
  grading of type $(14,7 )$:

{\small $$\begin{array}{l}
L_{  -1, -1, -1}=\langle b_{ 6}+b_{12 }-b_{ 16}-b_{18 }-b_{ 24}+b_{28 } ,\,- b_{ 10}+ b_{22 }\rangle\\
L_{-1, -1,
      1}=\langle b_{ 6}+b_{12 }+b_{ 16}+b_{18 }+b_{ 24}+b_{28 },\,  b_{ 10}+ b_{22 } \rangle
      \\ L_{-1,
      1, -1}=\langle -b_{ 11}+b_{13}+b_{ 15}-b_{23}+b_{ 25}+b_{27
      },\, b_{ 9}+ b_{21 }
      \rangle\\ L_{-1, 1, 1}=\langle  b_{ 11}+b_{13}-b_{ 15}-b_{23}-b_{ 25}+b_{27 },\,- b_{ 9}+ b_{21 } \rangle
      \\
      L_{1, -1, -1}=\langle -b_{ 5}+b_{8}-b_{ 14}+b_{17}-b_{ 20}+b_{26 },\,- b_{ 7}+ b_{19 } \rangle\\
      L_{1, -1,
      1}=\langle  b_{ 5}+b_{8}+b_{ 14}+b_{17}+b_{ 20}+b_{26 } ,\, b_{ 7}+ b_{19 } \rangle
      \\L_{1, 1, -1}=\langle b_{3 }, \,b_{ 1}+ b_{2} \rangle\\
      L_{-1, -1, \omega}=\langle \omega b_{ 6}+\omega^2  b_{ 12}+b_{16 }+\omega b_{ 18}+\omega^2  b_{ 24}+b_{ 28} \rangle\\
      L_{-1, -1, \
-\omega}=\langle \omega b_{ 6}+\omega^2  b_{ 12}-b_{16 }-\omega
b_{ 18}-\omega^2  b_{ 24}+b_{ 28} \rangle\\ L_{-1,
-1, -\omega^2  }=\langle \omega^2   b_{ 6}+\omega b_{ 12}-b_{16 }-\omega^2   b_{ 18}-\omega b_{ 24}+b_{ 28} \rangle\\
L_{-1, -1, \omega^2  }=\langle  \omega^2   b_{ 6}+\omega b_{ 12}+b_{16 }+\omega^2   b_{ 18}+\omega b_{ 24}+b_{ 28} \rangle\\
      L_{-1,
      1, \omega}=\langle  \omega^2   b_{ 11}+\omega b_{13 }-b_{15 }-\omega^2   b_{ 23}-\omega b_{ 25}+b_{27 } \rangle\\ L_{-1,
      1, -\omega}=\langle -\omega^2   b_{ 11}+\omega b_{13 }+b_{15 }-\omega^2   b_{ 23}+\omega b_{ 25}+b_{27 } \rangle\\ L_{-1,
      1, -\omega^2  }=\langle  -\omega b_{ 11}+\omega^2  b_{13 }+b_{15 }-\omega b_{ 23}+\omega^2  b_{ 25}+b_{27 } \rangle\\ L_{-1,
      1, \omega^2  }=\langle \omega b_{ 11}+\omega^2  b_{13 }-b_{15 }-\omega b_{ 23}-\omega^2  b_{ 25}+b_{27 } \rangle\\
      L_{1, -1, \omega}=\langle        \omega b_{5 }+\omega^2  b_{ 8}+b_{14 }+\omega b_{ 17}+\omega^2  b_{20 }+b_{26 } \rangle\\
      L_{      1, -1, -\omega}=\langle    -\omega b_{5 }+\omega^2  b_{ 8}-b_{14 }+\omega b_{ 17}-\omega^2  b_{20 }+b_{26 } \rangle\\
      L_{      1, -1, -\omega^2  }=\langle -\omega^2   b_{5 }+\omega b_{ 8}-b_{14 }+\omega^2   b_{ 17}-\omega b_{20 }+b_{26 } \rangle\\
      L_{      1, -1, \omega^2  }=\langle \omega^2   b_{5 }+\omega b_{ 8}+b_{14 }+\omega^2   b_{ 17}+\omega b_{20 }+b_{26 } \rangle\\
       L_{1,
      1, -\omega}=\langle \frac{-\omega-\omega^2  }{1-\omega}b_{1 }-\frac{2\omega}{1-\omega}b_{2 }-\frac{\omega}{1-\omega}b_{ 3}+ b_{4 } \rangle\\ L_{1,
      1, -\omega^2  }=\langle \frac{-\omega}{1-\omega}b_{1 }+\frac{2\omega^2  }{-1+\omega^2  }b_{2 }+\frac{\omega^2  }{-1+\omega^2  }b_{ 3}+ b_{4 } \rangle  \end{array}$$}

\end{document}